\documentclass[conference]{IEEEtran}
\IEEEoverridecommandlockouts

\usepackage{cite}
\usepackage{amsmath,amssymb,amsfonts}
\usepackage{algorithmic}
\usepackage{graphicx}
\usepackage{textcomp}
\usepackage{xcolor}
\def\BibTeX{{\rm B\kern-.05em{\sc i\kern-.025em b}\kern-.08em
    T\kern-.1667em\lower.7ex\hbox{E}\kern-.125emX}}
    
\usepackage[colorinlistoftodos]{todonotes}

\usepackage{atbegshi,picture}
\AtBeginShipoutNext{\AtBeginShipoutUpperLeft{%
  \put(\dimexpr\paperwidth-5.1cm\relax,-1.5cm){\makebox[0pt][r]{Accepted manuscript 11th International IEEE/EMBS Conference on Neural Engineering (NER) 2023}}%
}}

\begin{document}

\title{Towards AI-controlled FES-restoration of arm movements: neuromechanics-based reinforcement learning for 3-D reaching}

\author{Nat Wannawas$^1$ \&
        A. Aldo Faisal$^{1,2}$
\thanks{$^{1}$Brain \& Behaviour Lab, Imperial College London, London, United Kingdom. (email:nat.wannawas18@imperial.ac.uk). $^2$ Chair in Digital Health \& Data Science, University of Bayreuth, Bayreuth, Germany. (aldo.faisal@imperial.ac.uk).
NW acknowledges his support by the Royal Thai Government Scholarship.
AAF acknowledges his support by UKRI Turing AI Fellowship (EP/V025449/1).}
}


\maketitle

\begin{abstract}
    Reaching disabilities affect the quality of life. Functional Electrical Stimulation (FES) can restore lost motor functions. Yet, there remain challenges in controlling FES to induce desired movements. Neuromechanical models are valuable tools for developing FES control methods. However, focusing on the upper extremity areas, several existing models are either overly simplified or too computationally demanding for control purposes. Besides the model-related issues, finding a general method for governing the control rules for different tasks and subjects remains an engineering challenge.
    
    Here, we present our approach toward FES-based restoration of arm movements to address those fundamental issues in controlling FES. Firstly, we present our surface-FES-oriented neuromechanical models of human arms built using well-accepted, open-source software. The models are designed to capture significant dynamics in FES controls with minimal computational cost. Our models are customisable and can be used for testing different control methods. Secondly, we present the application of reinforcement learning (RL) as a general method for governing the control rules. In combination, our customisable models and RL-based control method open the possibility of delivering customised FES controls for different subjects and settings with minimal engineering intervention. We demonstrate our approach in planar and 3D settings.
\end{abstract}

\begin{figure*}[ht!]
    \begin{center}
    \includegraphics[width=0.92\linewidth]{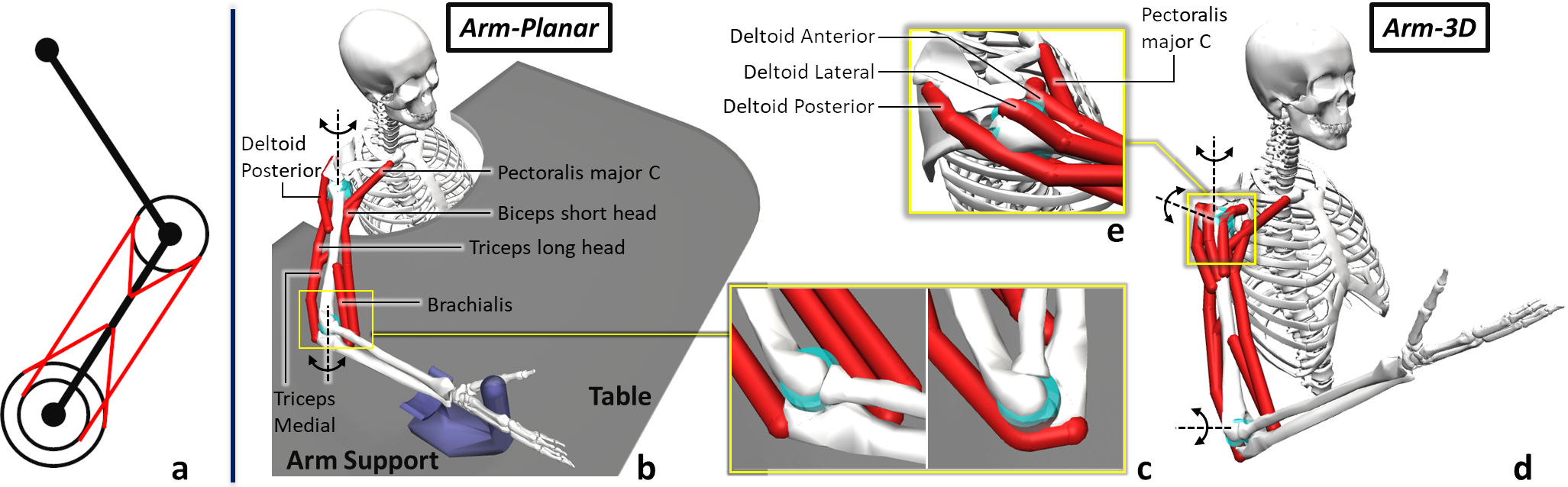}
    \vspace{-5pt}
    \caption{(a) An example of simple planar arm models. (b) Our neuromechanics \emph{Arm-Planar} model. (c) An illustration of muscle wrapping at the elbow. (d) Our neuromechanics \emph{Arm-3D} model and (e) its shoulder muscles.
    \label{fig:Models}}
    \end{center}
    \vspace{-15pt}
\end{figure*}

\section{Introduction}
Every year, stroke and spinal cord injury cause the loss of motor functions in individuals worldwide. In these cases, the limbs are technically functional but fail to receive motor commands from the brain. Arm movements are one of the commonly lost motor functions and cause severe limitations in performing daily tasks. Functional Electrical Stimulation (FES) or neuromuscular stimulation uses electrical signals to stimulate the muscle to induce its contraction and eventually movement of the limb. FES can be used to animate paralysed muscles and may help to restore nature motor functions in incomplete paralysis \cite{donaldson2000fes}. Early successes included work on lower limb paralysis, where the ability to cycling through rhythmic pedalling motions of the legs was restored. These successes beg the question how to extend this work to upper body restoration of movement. As FES cycling entails the periodic stimulation of muscles without the need for gravity compensation or end-point precision in their control, the movements are therefore relatively constrained and require relatively simple control. This is very different in arm movements where reaches towards an end-point require non-linear muscle coordination in the plane and gravity compensating activity in 3D movements in a volume. Early work in this nascent field therefore focus on single-joint control, e.g., of the elbow joint \cite{Wolf2020,Wannawas2022}. The literature on multiple-joint control of arm cases, however, is quite limited. This is partly because controlling FES in single-joint cases can be achieved through simple, model-free, error-based control such as PID controllers, while the multiple-joint cases require significantly more complex controls that have to include dynamical models in the systems.

Contrary to the fact that dynamical models play important roles in the controls and that the neuromechanics of the human arm is complex, many arm models used in FES control studies are relatively simple. In planar arm motion, for example, two-joint linkages with six muscles represented by straight lines models (Fig.\ref{fig:Models}a) are one of the most commonly used \cite{Izawa2004, Thomas2008, Jagodnik2016, Crowder2021, Abreu2022}. These models offer fast computation, but may not well capture the effects of muscle paths and their deformation during the movements. In addition, the muscles' properties themselves vary across the studies, thereby lacking standardisation and making the results difficult to reproduce or compare. On the other end, there exist commercial software such as \emph{LifeModeler} with highly detailed models. These models, however, are suitable for detailed analyses of particular situations such as ergonomic designs. Besides, the closed-source and commercial nature of the software could limit its usage among research communities, thereby not addressing the standardisation and reproducibility issues.

Besides the model-related issues, governing the control rules itself is a major challenge in inducing movements using FES. Regards specifically to inducing multiple-joint arm movements, successes in real-world settings are limited and, oftentimes, require assistive devices \cite{Freeman2015, Ambrosini2017, Wolf2020}. For example, the PID controller with inverse dynamics \cite{Wolf2020} can induce a narrow range of movements, while iterative learning control \cite{Freeman2015} can induce a longer range but is limited to repetitive trajectories. These limited successes are partly attributed to the difficulties of conventional methods in dealing with complexities and variations of human arms' neuromechanics. To our knowledge, an FES control method that can induce arbitrary movements across different subjects without intensive parameter tuning has not yet been reported.

Keeping both model-related and control governing issues in focus, we here present our approach towards FES-based restoration of arm movements that comprises two elements which, as separate entities, can address those issues. The first element is the neuromechanics models of the human arm built using OpenSim \cite{Delp2007}, a freely-available, open-source neuromechanical simulation software that is well-accepted in the communities. OpenSim allows us to build the models using established biomechanical components, e.g., muscle and joint models, thereby addressing the standardisation issue and providing state-of-the-art performances \cite{Saul2015}. Additionally, the open-source nature of OpenSim facilitate its uses for designing and testing different control methods which help promote reproducibility. In this work, we present our two arm models designed for surface FES control usages, i.e., they are designed for fast computation while maintaining important details. These models could be used as standards for comparing different control methods.

The second element of our approach is to govern the control policy using Reinforcement Learning (RL), a machine learning algorithm with a learning agent (RL agent) that learns to control an environment by interacting with it. RL can learn to control complex environments, for which hand-crafted control policies are difficult to govern. In FES control applications, RL is a promising method for governing the policies for arbitrary control settings. In addition, the fact that RL can learn customised stimulation for different subjects without intensive manual configuration can be an important factor that drives FES-based restoration of movements outside the laboratory and toward at-home usage. In this work, we present a generic RL setup to learn control policies for arbitrary arm-reaching tasks. We demonstrate the usage in planar and 3D reaching tasks using our OpenSim models.

\section{Related Works}
It is worth mentioning some related works to highlight their limitations and the gaps that this work fulfils. Regarding neuromechanical arm models, there exist several OpenSim models built by the communities. Closely related models are the OpenSim core \emph{Arm26} and \emph{MoBL-ARMS Dynamic Upper Limb} models \cite{Osimmodel}. These models have a few critical and minor issues as follows. The first critical issue is that they produce singularity computation in OpenSim4.4, the latest version, at some postures, causing crashes. Secondly, there is no mechanism such as joint limits to prevent unnatural postures. The minor issues are that there is no joint damping that prevents unnatural joint speed and, in some postures, the muscle paths are in the wrong positions, e.g., they wrap around the wrong side of the joint or go through the bone.

Regarding the applications of RL in FES control, the early studies were based on old RL algorithms, simple planar arm models (Fig.\ref{fig:Models}a), and a single, fixed target \cite{Izawa2004, Thomas2008, Jagodnik2016}. Recent studies have extended these settings to multiple targets \cite{Crowder2021,Abreu2022}. Our previous works investigate cycling motions in simulation \cite{Wannawas2021} and single-joint arm movements \cite{Wannawas2022} in the real world. A simulation study on 3D arm motions was conducted in \cite{Fischer2021} using a model that does not include the muscles, i.e., the RL agent directly controls the joint torques.

\section{Methods}
\paragraph{Neuromechanical models}
We use two human arm models; one is for planar motions (hereafter referred to as \emph{Arm-Planar}) which can be viewed as the detailed version of Fig.\ref{fig:Models}a-like model, and the other one is for 3D motions (hereafter referred to as \emph{Arm-3D}). Both models are designed at a suitable detail level for surface FES control applications, e.g., the muscles that are impossible to be stimulated separately via surface FES are bundled together to minimise the computation. The common properties and designs of both models are as follows. Both models have the right arms connected to the upper bodies located at fixed points in 3D space. The elbow and shoulder joints are modelled as pin and ball joints, respectively. Both joints have damping and joint limit mechanisms that prevent unnatural joint speed and postures. The muscles are built using a variant of Hill-type muscle model--\emph{DeGrooteFregly2016Muscle}. Both models have 4 muscles crossing elbows: \emph{Triceps Medial}, \emph{Triceps long head} (biarticular), \emph{Brachialis}, and \emph{Biceps short head} (biarticular). Note that \emph{Triceps lateral head} is bundled with \emph{Triceps long head}, and \emph{Biceps long head} is bundled with \emph{Biceps short head}. These muscles wrap around a cylindrical object at the elbows (Fig.\ref{fig:Models}c). The muscles' excitation-activation delay is changed from the default value of 40 ms to 100 ms to capture the long delay of FES-induced muscle activation \cite{Kralj1973}. The other muscle parameters such as maximum isometric force follow those in the \emph{Arm26} and \emph{MoBL-ARMS} models. The tendon slack length parameters are optimised using a genetic algorithm called CMAES \cite{Hansen2016} to equilibrate the passive forces. The other parts of both models have slightly different designs described as follows.

The \emph{Arm-Planar model} has 6 muscles in total. The other muscles besides the aforementioned 4 muscles are the \emph{Pectoralis major clavicular} head (\emph{Major C}) and \emph{Deltoid posterior}. These muscles wrap around a cylindrical object at the shoulder. The shoulder joint is only allowed to rotate around the vertical axis. The arm is supported under the wrist by an arm supporter (Fig.\ref{fig:Models}b) that moves on a table with low friction and provides gravity compensation to the arm. The \emph{Arm-3D model} has 8 muscles in total which are those 6 muscles of the \emph{Arm-Planar} model plus \emph{Deltoid lateral} and \emph{Deltoid anterior}. At the shoulder, there are 3 half ellipsoids functioning as muscle wrapper objects. These ellipsoids are carefully placed to support the full range of movements and prevent the wrong muscle path issues of the existing OpenSim models. The shoulder joint can rotate in all directions except the direction that causes the arm to twist.

\paragraph{Reinforcement Learning (RL) control}
 RL learns control policies through reward signals collected from the interaction with an environment. Here, interactions occur in a discrete-time fashion, starting with the agent observing the environment's state $\textbf{s}_t$ and selecting an action $\textbf{a}_t$ based on its policy $\pi$. The action causes the environment to be in a new state $\textbf{s}_{t+1}$. The agent then receives an immediate reward $r_t$ and observes $\textbf{s}_{t+1}$. This interaction experience is collected as a tuple $(\textbf{s}_t, \textbf{a}_t, r_t, \textbf{s}_{t+1})$ and stored in a replay buffer $\mathcal{D}$. This tuple is used to learn an optimal policy $\pi^*$ that maximises a return $R$--the sum of discounted immediate rewards.

The RL task here is to apply the muscle stimulation to move the arm to the desired pose which is specified by target joint angles--shoulder and elbow ($\boldsymbol{\theta}_{tar,t}$). The state vector $\textbf{s}_t$ is $[\boldsymbol{\theta}_t, \boldsymbol{\dot{\theta}}_t, \boldsymbol{\theta}_{tar,t}]^T$, where $\boldsymbol{\theta}_t$ and $\boldsymbol{\dot{\theta}}_t$ are the joint angles and angular velocities measured at time $t$, respectively. Note that appending the targets into the state vector allows the agents to learn goal-directed policies that can perform arbitrary reaching tasks. The action vector $\textbf{a}_t$ comprises normalised stimulation intensities ($i\in[0,1]$). The immediate reward $r_t$ is simply computed using the square error and action penalty as $r_t = -(\boldsymbol{\theta}_{t+1} - \boldsymbol{\theta}_{tar,t})^2 - (\Sigma_{i=0}^n a_i)/n$,
where $n$ is the number of stimulated muscles. With this setting, the optimal policy $\pi^*$ is simply the policy that causes the angles to be close to the targets with minimal stimulation.

The mechanism of finding the optimal policy varies across different RL algorithms. In this work, we choose the soft actor-critic (SAC) algorithm \cite{Haarnoja2019_1} because of its state-of-the-art performance in terms of both sample efficiency and stability across different environments. SAC has two components: an actor and a critic. In simple terms, the critic learns to estimate the expected return of a state-action pair, known as the Q value. The Q value is used to adjust the actor's policy $\pi$ by increasing the probability of choosing an action with a high Q value. Both actor and critic are parameterised by neural networks; we, based on empirical experiments and our previous works \cite{Wannawas2021,Wannawas2022}, use fully-connected neural networks that have two hidden layers. The output layer of the actor has a sigmoid activation function to squash the outputs.

The setups for the planar and 3D cases are slightly different. In the planar case, the involved angles are the elbow and shoulder angles which rotate about vertical axes. The state vector is therefore $\boldsymbol{s}\in\mathbb{R}^6$. The action vector $\boldsymbol{a}$ has 4 elements ($a_i\in[0,1]$) which are the normalised stimulation currents of \emph{Brachialis} and \emph{Biceps short head}, \emph{Triceps medial} and \emph{Triceps long head}, \emph{Pectoralis major c}, and \emph{Deltoid posterior}. We set Biceps stimulation to affect two muscles because, in real situations, only a single pair of electrodes are placed above \emph{Biceps} (and similarly for \emph{Triceps}). In the 3D case, the shoulder joint rotates in 2 directions, and the \emph{Deltoid lateral} and \emph{Deltoid anterior} are stimulated via the same pair of electrodes. Hence, the state vector becomes $\boldsymbol{s}\in\mathbb{R}^9$, and the action vector has 5 elements.

The RL training is episodic. Each episode starts with a random arm pose and target. Each episode has 100 time steps with 100 ms time-step size. The target changes to a new random value at the 50\textsuperscript{th} time step. Every 5 training episodes, the agents' performances are evaluated on 50 test episodes.

\vspace{-2pt}

\section{Results}
\begin{figure*}
    \begin{center}
    \includegraphics[width=0.975\textwidth]{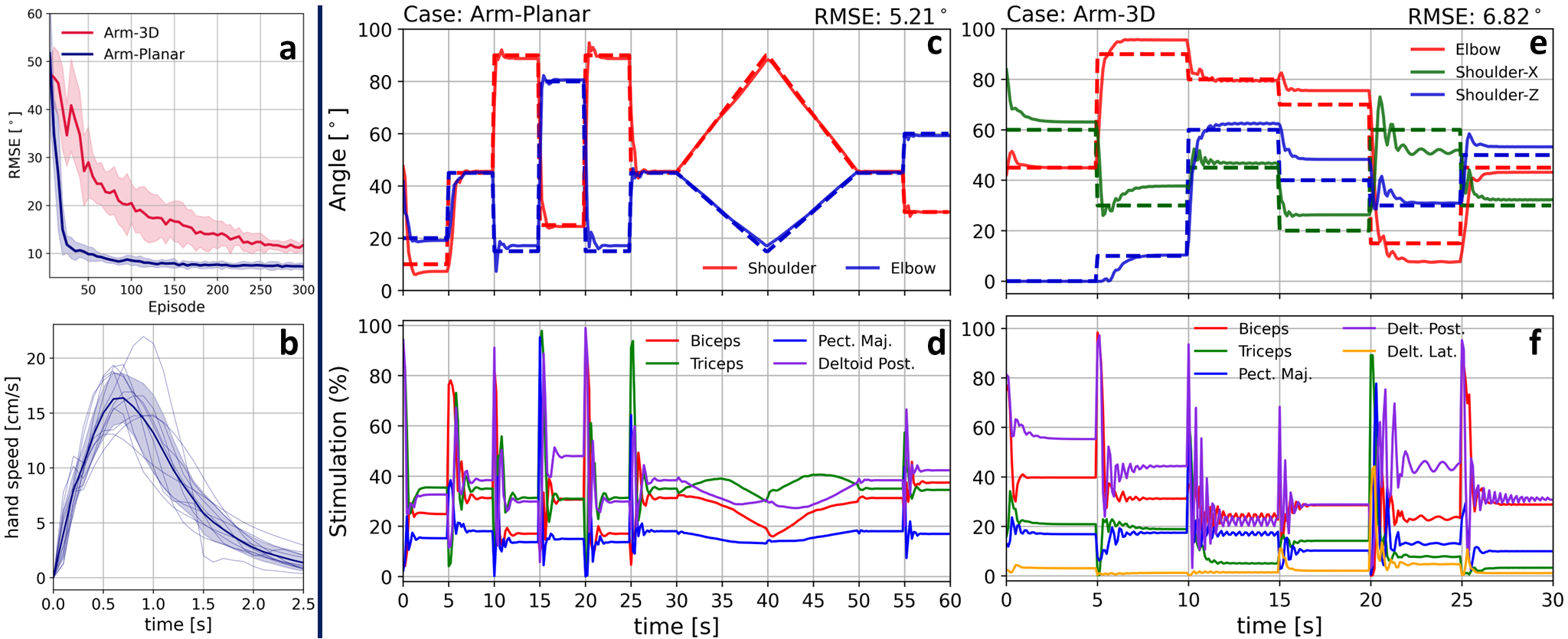}
    \vspace{-5pt}
    \caption{(a) Performance evaluation along the training in (\textcolor{red}{red}) Arm-3D and (\textcolor{blue}{blue}) Arm-Planar cases. The solid lines and the shades show the mean and standard deviation of 10 runs. (b) Hand speed curves along planar, point-to-point movements. The thin lines represent the curves induced by the individual RL agent. The thick line and shade represent the mean and standard deviation. The examples of trajectory tracking in (c) Arm-Planar and (e) Arm-3D cases. The dash and solid lines are the targets and actual angles that the RL agents achieve, respectively. (d) and (f) show the stimulation along the tracking.
    \label{fig:Results}}
    \end{center}
    \vspace{-20pt}
\end{figure*}

In both cases, RL agents are trained for 300 episodes. The training is repeated 10 times to evaluate the robustness. The performance evaluations along the training are shown in Fig.\ref{fig:Results}a. The average RMSE performance at the end of the training are approximately $7^\circ$ in the planar and $10^\circ$ in the 3D cases. Note that these RMSE values are computed over the entire movement trajectory, which is stricter measure than previous work that looks only at the end-point precision of movement. We therefore compute our worst-case end-point precision which is 5cm (all reaches within target that size). Our performance levels are comparable to those in \cite{Crowder2021,Abreu2022} where targets with the radius size of 5cm are accurately reached. The learning curves of the system in the planar case are significantly quicker than in the 3D case, and the variability across 10 trials are on the order of $2^\circ$ ranges suggesting how robust our method is. Fig.\ref{fig:Results}b shows hand velocity curves of planar, point-to-point movements. The movement trajectories look qualitatively natural, and crucially movements show the characteristic smooth, bell-shaped velocity profiles of natural reaching movements \cite{Burdet1998}.

Fig.\ref{fig:Results}c and e show examples of control performances in planar and 3D cases, respectively. In both cases, the RL agents can track arbitrary trajectories that have never been assigned during the training. The performance in the planar case is slightly better than that in the 3D case as the planar movements are less complex. Fig.\ref{fig:Results}d and f show the stimulation applied during the tracking tasks. In both cases, brief bursts of stimulation appear when the targets change, followed by steady stimulation that co-contraction the muscles to stabilise the arms.

\section{Discussion \& Conclusion}
We present our approach towards FES-based restoration of arm movements. The first element of our approach is the OpenSim-based modelling which helps facilitate the process and standardise the models on which different control methods are tested and compared. We also present our two OpenSim models: \emph{Arm-Planar} and \emph{Arm-3D}. The second element is the RL-based control governing which can provide customised stimulation for different subjects and control settings with minimal technical intervention. We present a generic RL training setup and demonstrate its applications on our OpenSim models in arbitrary reaching tasks.

Several further steps might be explored to translate this method into real usages. One step is to optimise the models to accurately represent the dynamics of a certain subject's arm. The customised model can be used for pre-training the RL before transferring it to the real subject. Another step is to take muscle fatigue into account. The fatigue will cause the environment's state to become partially observable, causing problems in typical RL's learning mechanism. Our sister paper \cite{Wannawas2023b} presents a machine-learning based extension that allows RL to learn to compensate for fatigue dynamically, enabling continuous training in practical situations.


In conclusion, we show how the integration of reinforcement learning methods can be combined with detailed neuromechanical models to address existing challenges in FES control. There is thus hope that FES-based restoration of interactive movements in paralysed people can exceed the capabilities of recent advances in assistive technologies \cite{Shafti2019}.

\bibliographystyle{IEEEtran}
\bibliography{bib}

\begin{thebibliography}{10}
\providecommand{\url}[1]{#1}
\csname url@samestyle\endcsname
\providecommand{\newblock}{\relax}
\providecommand{\bibinfo}[2]{#2}
\providecommand{\BIBentrySTDinterwordspacing}{\spaceskip=0pt\relax}
\providecommand{\BIBentryALTinterwordstretchfactor}{4}
\providecommand{\BIBentryALTinterwordspacing}{\spaceskip=\fontdimen2\font plus
\BIBentryALTinterwordstretchfactor\fontdimen3\font minus
  \fontdimen4\font\relax}
\providecommand{\BIBforeignlanguage}[2]{{%
\expandafter\ifx\csname l@#1\endcsname\relax
\typeout{** WARNING: IEEEtran.bst: No hyphenation pattern has been}%
\typeout{** loaded for the language `#1'. Using the pattern for}%
\typeout{** the default language instead.}%
\else
\language=\csname l@#1\endcsname
\fi
#2}}
\providecommand{\BIBdecl}{\relax}
\BIBdecl

\bibitem{donaldson2000fes}
N.~Donaldson \emph{et~al.}, ``Fes cycling may promote recovery of leg function
  after incomplete spinal cord injury,'' \emph{Spinal Cord}, vol.~38, no.~11,
  pp. 680--682, 2000.

\bibitem{Wolf2020}
D.~N. Wolf, Z.~A. Hall, and E.~M. Schearer, ``Model learning for control of a
  paralyzed human arm with functional electrical stimulation,'' in \emph{IEEE
  Intl Conf Robotics and Automation (ICRA)}, 2020.

\bibitem{Wannawas2022}
N.~Wannawas, A.~Shafti, and A.~A. Faisal, ``Neuromuscular reinforcement
  learning to actuate human limbs through fes,'' in \emph{IFESS22}, 2022.

\bibitem{Izawa2004}
J.~Izawa \emph{et~al.}, ``Biological arm motion through reinforcement
  learning,'' \emph{Biol. Cyber.}, vol.~91, pp. 10--22, 2004.

\bibitem{Thomas2008}
P.~Thomas \emph{et~al.}, ``Creating a reinforcement learning controller for
  functional electrical stimulation of a human arm,'' in \emph{14th Yale
  Workshop on Adaptive and Learning Systems}, 2008.

\bibitem{Jagodnik2016}
K.~M. Jagodnik \emph{et~al.}, ``Human-like rewards to train a reinforcement
  learning controller for planar arm movement,'' \emph{IEEE Trans on
  Human-Machine Systems}, vol.~46, pp. 723--733, 2016.

\bibitem{Crowder2021}
D.~C. Crowder, J.~Abreu, and R.~F. Kirsch, ``Hindsight experience replay
  improves reinforcement learning for control of a mimo musculoskeletal model
  of the human arm,'' \emph{IEEE Trans on Neural Sys. \& Rehabil. Eng.},
  vol.~29, pp. 1016--1025, 2021.

\bibitem{Abreu2022}
J.~Abreu \emph{et~al.}, ``Deep reinforcement learning for control of
  time-varying musculoskeletal systems with high fatigability: a feasibility
  study,'' in \emph{IEEE Trans. Neural Sys. and Rehab. Eng.}, 2022.

\bibitem{Freeman2015}
C.~T. Freeman, ``Upper limb electrical stimulation using input-output
  linearization and iterative learning control,'' \emph{IEEE Trans. Control
  Systems Technology}, vol.~23, pp. 1546--1554, 2015.

\bibitem{Ambrosini2017}
E.~Ambrosini \emph{et~al.}, ``The combined action of a passive exoskeleton and
  an emg-controlled neuroprosthesis for upper limb stroke rehabilitation: First
  results of the retrainer project,'' in \emph{IEEE Intl Conf Rehabil Robot.
  (ICORR)}, 2017, pp. 56--61.

\bibitem{Delp2007}
S.~L. Delp \emph{et~al.}, ``Opensim: Open-source software to create and analyze
  dynamic simulations of movement,'' \emph{IEEE Trans. Biomed. Eng.}, vol.~54,
  pp. 1940--1950, 2007.

\bibitem{Saul2015}
K.~R. Saul \emph{et~al.}, ``Benchmarking of dynamic simulation predictions in
  two software platforms using an upper limb musculoskeletal model,''
  \emph{Computer Methods in Biomec. \& Biomed. Eng.}, vol.~18, pp. 1445--1458,
  2015.

\bibitem{Osimmodel}
\BIBentryALTinterwordspacing
Opensim documentation. [Online]. Available:
  \url{https://simtk-confluence.stanford.edu:8443/display/OpenSim/Musculoskeletal+Models}
\BIBentrySTDinterwordspacing

\bibitem{Wannawas2021}
N.~Wannawas, M.~Subramanian, and A.~A. Faisal, ``Neuromechanics-based deep
  reinforcement learning of neurostimulation control in fes cycling,''
  \emph{IEEE/EMBS Neural Eng. (NER)}, vol.~10, pp. 381--384, 2021.

\bibitem{Fischer2021}
F.~Fischer \emph{et~al.}, ``Reinforcement learning control of a biomechanical
  model of the upper extremity,'' \emph{Scientific Reports}, vol.~11, 2021.

\bibitem{Kralj1973}
A.~Kralj and S.~Grobelnik, ``Functional electrical stimulation - a new hope for
  paraplegic patients?'' \emph{Bulletin of Prosthetics Research, University of
  Ljubljana, Yugoslavia}, pp. 75--102, 1973.

\bibitem{Hansen2016}
N.~Hansen, ``The cma evolution strategy: A tutorial,''
  \emph{arXiv:1604.007722v1 [cs.LG]}, 2016.

\bibitem{Haarnoja2019_1}
T.~Haarnoja \emph{et~al.}, ``Soft actor-critic algorithms and applications,''
  \emph{arXiv:1812.05905v2 [cs.LG]}, 2019.

\bibitem{Burdet1998}
E.~Burdet and T.~Milner, ``Quantization of human motions and learning of
  accurate movements,'' \emph{Biol Cybern.}, vol.~78, pp. 307--318, 1998.

\bibitem{Wannawas2023b}
N.~Wannawas and A.~A. Faisal, ``Towards ai-controlled fes-restoration of arm
  movements: Controlling for progressive muscular fatigue with gaussian
  state-space models,'' in \emph{IEEE/EMBS Neural Eng.}, 2023.

\bibitem{Shafti2019}
A.~Shafti, P.~Orlov, and A.~A. Faisal, ``Gaze-based, context-aware robotic
  system for assisted reaching and grasping,'' in \emph{Intl. Conf. on Robotics
  and Automation (ICRA)}, 2019.

\end{thebibliography}

\end{document}